\documentclass[twocolumn,prb,aps,showpacs]{revtex4}
\usepackage{graphics}

\usepackage{dcolumn}

\usepackage{bm}

\usepackage{epsfig}

\begin{document}

\title{Simple model for scanning tunneling spectroscopy
of noble metal surfaces with adsorbed Kondo impurities}
\author{J. Merino and O. Gunnarsson}
\affiliation{Max-Planck-Institut f\"ur Festk\"orperforschung
D-70506 Stuttgart, Germany}
\date{\today}
\begin{abstract}
A simple model is introduced to describe conductance measurements
between a scanning tunneling microscope (STM) tip and a noble
metal surface with adsorbed transition metal atoms which display the 
Kondo effect.  The model
assumes a realistic parameterization of the potential created by
the surface and a $d_{3z^2-r^2}$ orbital for the description of
the adsorbate. Fano lineshapes associated with the Kondo resonance 
are found to be sensitive to
details of the adsorbate-substrate interaction.  For instance,
bringing the adsorbate closer to the surface leads to more
asymmetric lineshapes while their dependence on the tip distance
is weak. We find that it is important to use a realistic surface
potential,  to properly include the tunnelling matrix elements
to the tip and to use substrate states which are orthogonal to the 
adsorbate and tip states.  An application of our model to Co adsorbed on Cu
explains the difference in the lineshapes observed between Cu(100)
and Cu(111) surfaces.
\end{abstract}

\maketitle

\section{Introduction}

The study of many-body phenomena in low dimensional systems is
attracting a lot of attention.  This has been motivated by recent
advances in the construction of nanostructures and quantum dot
devices. Scanning tunneling microscopy has also opened the
possibility of analyzing many-body phenomena at surfaces. For
instance, the Kondo effect has been detected in STM conductance
measurements of noble metal surfaces with adsorbed 3d transition
metal atoms.\cite{Madhavan1,Manoharan,Knorr,Nagaoka,Li} Characteristic zero bias lineshapes
are observed which are reminiscent of Fano\cite{Fano} phenomena.

Fano lineshapes have been observed in different situations. They
were first explored by Fano \cite{Fano} in his studies of
autoionization of doubly excited He(2s2p) resonances lying in the
continuum.  In the case of adsorbed atoms on a metal surface an
analogous situation is found as a localized orbital is also
coupled to a continuum of metallic states. As the adsorbate is
magnetic,  the Kondo effect can occur and, therefore, Fano
lineshapes can be thought of as arising from the interference of the
Kondo resonance with the continuum of metal states.\cite{Ujsaghy}
It is worth noting that Fano phenomena appears in STM measurements
although the tip overlaps much more strongly with the surface than
with the adsorbate wavefunctions (as the 3d orbitals are very
localized) so that conductance measurements reflect electronic
properties of the metal surface modified by the presence of the
magnetic atom.

These Fano-type lineshapes differ from one adsorbate/substrate
system to another, as summarized in Table \ref{table1}, although
some experimental trends can be extracted.  For instance, the Fano
parameter, $q$, is  typically either zero or positive.  As can be
observed from the table the lineshape associated with $q=0$ is a
symmetric dip close to zero bias.  The dependence of lineshapes
may be illustrated by comparing the asymmetric dip observed for Co
on Cu(100) with the symmetric dip-like shape found for Co on
Cu(111). Similarly adsorbing Ti instead of Co on Au(111) leads to a strong
variation of the lineshape. The situation becomes more complicated
if we consider the "middle" elements of the 3d row which do not
even show appreciable features in the conductance\cite{Jamneala}
down to $T=6$ K. This raises questions about the occurrence of the
Kondo effect at all for these specific adsorbates. Hence,
experimental observations suggest that details associated with the
adsorbate-substrate interaction may be relevant.
\begin{table}
\caption{Non-universality of lineshapes observed in conductance
measurements between a STM tip and a noble metal surface with
adsorbed 3d transition metal atoms and Ce.  Depending on the
adsorbate, the type of noble metal and/or the surface face,
observed conductance lineshapes are different.  The Fano
parameter, $q$, which measures the degree of asymmetry of the
lineshape associated with the Kondo resonance is typically $q \geq
0$. T$_K$ denotes the associated Kondo temperatures given in
degrees Kelvin.  } \label{table1}
\begin{tabular}{lllll}
Adsorbate/Surface & T$_K$(K) & Type of lineshape &
q & Ref.\\
\hline
Co/Cu(111) &  54 & nearly symmetric dip & 0.2 & [\onlinecite{Manoharan,Knorr}] \\
Co/Cu(100) &  88 & asymmetric dip &  1.1 & [\onlinecite{Knorr}] \\
Co/Au(111) &  75 & asymmetric dip  & 0.6 & [\onlinecite{Madhavan1}]\\
Ti/Au(111) &  70 & asymmetric dip+peak &  1  & [\onlinecite{Jamneala}]\\
Ti/Ag(100) &  40 & asymmetric dip+peak  &  1  & [\onlinecite{Nagaoka}] \\
Ce/Ag(111) &  500 & symmetric dip  &  0 & [\onlinecite{Li}]\\
Co/Ag(111) &  92 & symmetric dip & 0 & [\onlinecite{Schneider}]\\ \hline
\end{tabular}
\end{table}

The Fano parameter, $q$, governing the shape of the conductance,
$G(\omega)$, close to zero bias, is given by:
\begin{equation}
q= {A(\epsilon_F) \over B(\epsilon_F) },
\label{fanoparam}
\end{equation}
where $B(\omega)$ reads:
\begin{equation}
B(\omega) = \pi \sum_{\bf k}
M_{ \bf k} V_{\bf k}\delta(\omega -\epsilon_{\bf k}),
\label{B}
\end{equation}
and, $A(\omega)$, is the Kramers-Kronig transformation of $B(\omega)$:
\begin{equation}
A(\omega) = {1 \over \pi} \int_{-\infty}^{\infty}
d \omega' {B(\omega') \over \omega-\omega'}.
\label{A}
\end{equation}
In the above equations, $M_{\bf k}$ and  $V_{\bf k}$ are the tip-substrate and
adsorbate-substrate matrix elements, respectively. $\epsilon_{\bf k}$ is
the substrate band dispersion and $\epsilon_F$ the position of the Fermi level.

In order to understand the relation between the energy dependence of
$B(\omega)$ and the conductance lineshape we consider first in Fig. \ref{fig1}
two different situations attending to the degree of asymmetry of
$B(\omega)$:
\begin{figure}
\begin{center}
\epsfig{file=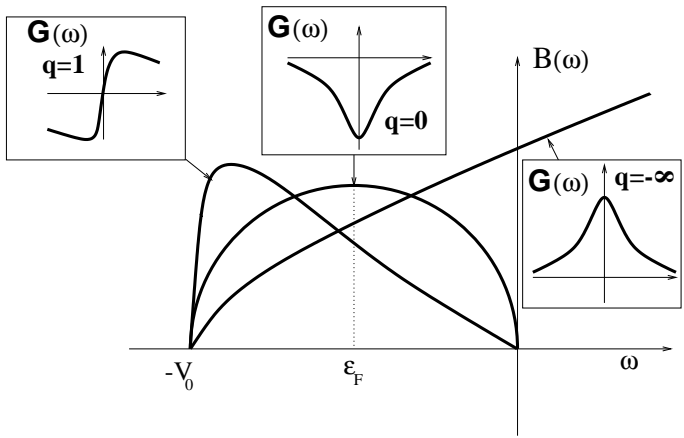,width=9.0cm,angle=0}
\end{center}
\caption{Correspondence of shape of $B(\omega)$ with the Fano
parameter. This schematic plot gives a qualitative understanding
of how the shape of the conductance, $G(\omega)$, around the
Fermi energy is related to $B(\omega)$.  The Fermi energy refered to the
vacuum level is denoted by the dashed vertical line $\epsilon_F=-W$,
where $W$ is the metal workfunction. $-V_0$ denotes the bottom of the band.} \label{fig1}
\end{figure}
(i) $B(\omega)$ is symmetric with respect to the Fermi energy. In
this case, $A(\epsilon_F)=0$ and the Fano parameter, $q=0$. Hence,
the resulting conductance lineshape is a symmetric dip. (ii)
$B(\omega)$ is asymmetric with most of its weight situated at
frequencies below the Fermi energy $\omega < \epsilon_F$. For this
case, $q>0$ and conductance lineshapes are more asymmetric 
($q = 1$ in the most asymmetric lineshape).  Hence, experimental
observations requires that $B(\omega)$ is in between cases (i) and
(ii).

However, if we consider an sp-like band for the substrate and assume that hybridization matrix
elements are independent of momentum, $B(\omega)$ would be
parabolic (from Eq.~(\ref{B})),  growing up to infinite energies.  Due to the large asymmetry in
$B(\omega)$, $A(\epsilon_F)$ would be very large and negative
so that $q \rightarrow -\infty$ leading to a peak (see Fig. \ref{fig1}) rather than the dip typically
observed in experiments. Hence, it seems difficult to reconcile the observed conductance
lineshapes with the shape of $B(\omega)$ from this simplified picture. The
question that we address in the present work is whether the momentum dependence of 
hybridization matrix elements can change $B(\omega)$ leading to more
symmetric conductance lineshapes. 

First attempts to model the substrate electronic structure have been carried
out by Plihal and Gadzuk\cite{Plihal} who have used a Jellium
metal with a sharp step potential barrier at the
surface.  However, a half-filled symmetric density of states to describe the substrate
(instead of parabolic) and momentum independent adsorbate-substrate hybridization matrix elements, $V_{\bf k}$,
were used. Under these strong assumptions dip-like conductance lineshapes were obtained.
Tight-binding descriptions of the substrate have also been used.\cite{Schiller}
They provide a qualitative understanding of observations although their use is difficult
to justify considering the sp free-like bands of noble metal surfaces.

The above discussion shows the difficulty of describing
experimental observations in a consistent way and points out
the need of a more realistic model for describing the adsorbate-substrate-tip
system.

An issue which needs to be carefully addressed in defining the
relevant model is to know which of the states associated
with the surface (either surface and/or bulk states) are more
strongly coupled to the adsorbate.  There is experimental evidence
suggesting that surface states play only a minor role in the
metal-adsorbate interaction \cite{Knorr}.  For instance, the amplitude of the
conductance decays rapidly as, $G \propto 1/R_{||}^2$, with
lateral displacement of the tip,  $R_{||}$,  instead of decaying
as $1/R_{||}$ expected for surface states coupled to the impurity.
Furthermore, this bulk-like behavior persists in conductance
measurements of Co on Cu(111), which is known to have a surface
state at the Fermi energy \cite{Knorr}. Based on the above
experimental observations we will only consider bulk states in our
model.  Another experimental observation is that lineshapes
depend only weakly with perpendicular tip-substrate distance, $Z_t$,
suggesting that the direct interaction between the tip and the
adsorbate is negligible\cite{Madhavan2}.

In the present work we show how using an Anderson model in an
appropriate orthogonalized basis, reasonably symmetric shapes of
$B(\omega)$ are found which lead to conductance lineshapes in
agreement with observations. We find that the momentum dependence
of the hybridization matrix elements between the adsorbate
$3d_{3z^2-r^2}$ orbital and the orthogonalized metal wavefunctions together with
the finite size of the tip wavefunction are responsible for this
behavior.

The present paper is organized as follows.  In Section
\ref{sectheo} we introduce an Anderson model in an orthogonalized
basis to describe the adsorbate-substrate interaction together
with the relevant formulas needed for discussing STM conductance
measurements. In Section \ref{secdis} we compute the parameters
involved in the adsorbate- substrate interaction needed in the
Anderson model.  Section \ref{secres} is devoted to describing the
main results of our model.  Finally,  in Section \ref{secco}, we
apply the model proposed to conductance measurements of Co atoms
on noble metal surfaces.

\section{Theoretical approach}

\label{sectheo}

In this section, details about the model used are given.  An Anderson model on
an orthogonal basis
is considered.  The parameters of the model are computed taking into account the
following considerations:
(i) Instead of a step barrier we consider
a Jones-Jennings-Jepsen (JJJ) potential \cite{Jones} to describe the metal surface
wavefunctions,
(ii) we neglect the direct coupling of the tip with the substrate d bands
and with the 3d orbital of the adsorbate due to the localized nature of
the d orbitals,
(iii) the adsorbate is modelled by a single d-orbital, and
(iv) the momentum dependence of the
hybridization matrix elements is explicitly taken into account.

Metallic states, $|{\bf k}>$, coupled to a single
$d_{3z^2-2}$ orbital denoted by $|d>$ are considered.
The Anderson model usually assumes that the cotinuum of metal
states are orthogonal to the localized orbitals of the adsorbate.  However, the
basis set
formed by the unperturbed metal, adsorbate and tip wavefunctions,
$\{ |{\bf k}>, |d>, |t> \}$, is non-orthogonal (in general) and overcomplete.

One way to take into account orthogonalization effects
is to redefine the metallic states, ${\bf k}$, as:
\begin{equation}
|{\bf \tilde{k} }> = |{\bf k}> - <d|{\bf k}>|d>-<t|{\bf k}>|t>.
\label{ortk}
\end{equation}
Considering that tip and adsorbate wavefunctions are
orthogonal: $<t|d>=0$, as $|d>$ is very localized, then the new metallic
states satisfy
\begin{equation}
<\tilde{{\bf k}}|\phi>=0,
\end{equation}
where $|\phi>$ can be either $|t>$ or $|d>$.

Our starting point is an Anderson model defined in this new
orthogonal basis: $\{ |{\bf \tilde{k}}>, |d>, |t> \}$, from which
associated one-electron parameters, $\epsilon_{\bf \tilde{k}}$,
$V_{\bf \tilde{k}}$ and $M_{\bf \tilde{k}}$ are obtained.  An
analogous procedure was previously used in the context of
chemisorption of atoms and molecules on metal surfaces by Grimley.
\cite{Grimley}

It is worth mentioning that the new metallic states are
non-orthogonal among each other:
\begin{equation}
<\tilde{{\bf k}}|\tilde{{\bf k'}}> \neq 0.
\end{equation}
Hence, the orthogonalization condition between different
$|\tilde{{\bf k}}>$ states is violated. However,  this occurs at
higher order in the overlap: $O(<{\bf k}|d>^2)$.

\subsection{Model}
From the above considerations, our model for the complete tip-substrate system
(with the 3d transition metal atom) reads
\begin{equation}
H=H_{subs}+H_{tip-subs}+H_{tip},
\end{equation}
where $H_{subs}$ describes the substrate with the adsorbed
3d transition metal atom and $H_{tip-subs}$ describes the
interaction of the tip with the substrate.  $H_{tip}$ describes
the tip which is assumed to have an unstructured density of states.

The substrate with the adsorbed 3d atom may be
modelled by a generalized Anderson model that
explicitly includes the orbital degeneracy of the 3d-orbital
\begin{eqnarray}
H_{subs}&=&\sum_{{\bf k},m,\sigma } \epsilon_{\bf \tilde{k}
} c^{\dagger}_{{\bf \tilde{k}}m\sigma} c_{{\bf \tilde{k}}m\sigma} +
\epsilon_d \sum_{m,\sigma} d^{\dagger}_{m\sigma} d_{m\sigma}
\nonumber  \\
&+& \sum_{ {\bf k},m,\sigma} V_{{\bf \tilde{k}}}
(d^{\dagger}_{m\sigma} c_{{\bf \tilde{k}}m\sigma} + H. c.)
\nonumber \\
&+&U \sum_{(m,\sigma) < (m',\sigma')} d^{\dagger}_{m\sigma}
d_{m,\sigma} d^{\dagger}_{m'\sigma'} d_{m'\sigma'}.
\label{subs}
\end{eqnarray}
Here $\epsilon_d$ is the energy level of an electron residing in
the $d$ orbital of the adsorbate, $c^{\dagger}_{{\bf \tilde{k}}m
\sigma}$ creates an electron with spin $\sigma$, momentum ${\bf
\tilde{k}}$ and perpendicular projection of angular momentum, $m$,
in the metal. $d^{\dagger}_{m\sigma}$ creates an electron in the
state with perpendicular projection of angular momentum $m$ in the
adsorbate. $\epsilon_{\bf \tilde{k}}$ and  $V_{{\bf \tilde{k}}}$
are the metallic energies and the hybridization matrix elements
between the substrate and the adsorbate, respectively. $U$ is the
Coulomb repulsion of two electrons in the 3d orbital of the
transition metal atom.

We will restrict the sum to $m=0$
as this is the orbital which is more strongly hybridized to the
metallic surface.\cite{Weissman}.  In this case, Hamiltonian~(\ref{subs})
reduces to the standard Anderson impurity model containing the spin
degeneracy only.

Finally, the tip-substrate interaction contribution to the
Hamiltonian reads
\begin{equation}
H_{tip-subs}=\sum_{{\bf k},\sigma} M_{ {\bf \tilde{k}}}
( c^{\dagger}_{{\bf \tilde{k}}\sigma } t_{\sigma} + H. c.),
\label{tip}
\end{equation}
through the matrix elements, $M_{{\bf \tilde{k}}}$. Here,
$t_{\sigma}$ destroys an electron with spin $\sigma$
in the tip.

\subsection{Hybridization matrix elements}
In the following we describe how hybridization matrix elements,
$M_{{\bf \tilde {k}}}$ and $V_{{\bf \tilde{k}}}$, are computed.
For simplicity we will first focus on how $V_{\bf \tilde{k}}$ is
computed, as $M_{{\bf \tilde{k}}}$ is computed in a similar way.

Ignoring the electron-electron interaction we can re-express Hamiltonian~(\ref {subs})
in first quantized form
\begin{equation}
H_{subs}=T+V_d+V_M,
\label{first}
\end{equation}
where $T$ is the kinetic energy of the system, $V_d$ is the
potential created by the adsorbate and $V_M$ describes the
surface potential.

From Eqs.~(\ref{ortk}) and (\ref{first}), the matrix elements between the
orthogonalized metallic states, ${\bf \tilde{k}}$, and the adsorbate
read:
\begin{equation}
V_{\bf \tilde{k}}= V_{\bf k}-S_{\bf k}<d|V_M|d>,
\label{matrixel}
\end{equation}
where, again, we have assumed: $<t|d>=0$.

The first term in Eq.~(\ref{matrixel}) is
the hybridization matrix element with the
unperturbed wavefunctions ${\bf k}$:
\begin{equation}
V_{\bf k} = <{\bf k}|V_M|d>,
\end{equation}
and the second contains the overlap matrix element
\begin{equation}
S_{\bf k} = <{\bf k}|d>.
\end{equation}

The above orthogonalization procedure
automatically selects the metal potential $V_M$ in the
hybridization matrix elements favouring the region close to or inside the metal
in the integrations. This differs from hybridization matrix elements
computed with the original wavefunctions as in that case integrations
over the whole space are involved.

Finally we note that orthogonalization effects enter the
model through matrix elements only. Orthogonalization effects on
the substrate band energies can be shown to be of higher order in the
overlap. Hence, we will assume $\epsilon_{\bf \tilde{k}}=\epsilon_{\bf k}$ in the
rest of the paper.

\subsection{Computation of conductance}
Following Ref.~[\onlinecite{Madhavan2}] and for the sake of clarity
we derive the basic equations needed for the computation of the
conductance through the STM. If we neglect any modification of the
substrate due to the presence of the tip (this is reasonable
considering the fact that the tip is typically at about $5-10$ \AA~above the metal surface), 
then the conductance measured by the STM
reads: \cite{Madhavan2,Plihal,Schiller}
\begin{equation}
G(\omega) = {4 e^2 \over \hbar} \rho_{tip}
(\Gamma(\omega) + \delta \Gamma(\omega)),
\label{conduct}
\end{equation}
where
\begin{equation}
\Gamma(\omega) = \pi \sum_{\bf k}|M_{{\bf \tilde{k}}}|^2
\delta(\omega-\epsilon_{\bf k})
\end{equation}
is the conductance associated with the clean substrate (without the
adsorbed 3d transition metal atom) and $\rho_{tip}$ is the density of states of
the tip.

Modifications of the tip-surface coupling induced by the presence of the
adsorbate are given by
\begin{equation}
\delta \Gamma(\omega)= {\rm Im} \sum_{{\bf k, k'}}
{M_{ {\bf \tilde{k} } } V_{ {\bf \tilde{k} } } \over \omega - \epsilon_{\bf k} -
i \eta}
G_{dd}(\omega)
{M^*_{ {\bf \tilde{k'}} } V^*_{ {\bf \tilde{k'}} } \over \omega - \epsilon_{\bf
k'} -i \eta}.
\label{gamma}
\end{equation}

In Eq.~(\ref{conduct}), $\rho_{tip}$ is the density of states of the tip.
$G_{dd}(\omega)$ describes the
electronic properties of the $3d$ adsorbate immersed in the
metallic continuum including the many-body effects such as the
Kondo effect induced by the on-site Coulomb interaction, $U$, inside
the localized $3d$ orbital.  $\eta$ is an analytical continuation parameter.

For convenience Eq.~(\ref{gamma}) is rewritten in the following way
\begin{equation}
\delta \Gamma(\omega) =
{\rm Im} \{ (A(\omega) + i B(\omega)) G_{dd}(\omega)
(A^*(\omega)+i B^*(\omega)) \}
\label{gammaAB}
\end{equation}
where $B(\omega)$ is defined in Eq.~(\ref{B}) and $A(\omega)$ is the
Kramers-Kronig transformation of $B(\omega)$ given in Eq.~(\ref{A}) with
the matrix elements evaluated with the orthogonalized wavefunctions,
$|{\bf \tilde{k}}>$.  For the systems of interest here, $A(\omega)$
and $B(\omega)$ are real.  Function $B(\omega)$ embodies the information concerning the
tip-substrate-adsorbate system as it depends on the tip-adsorbate
separation, ${\bf R}$, the position of the adsorbate with respect
to the plane of ions, $Z_d$, and the metal potential described by
$V_M$ through the matrix elements, $V_{\bf \tilde{k}}$ and $M_{\bf
\tilde{k}}$. In the present work we analyze how $B(\omega)$ and
$\delta \Gamma (\omega)$ depend on these parameters.

From Eq.~(\ref{gammaAB}) we notice that the
conductance can, in principle, have
any kind of shape as a result of the interference
of the adsorbate with the substrate continuum of states.  In the
following we will see how, in fact, Eq.~(\ref{conduct}) reduces
to the well known Fano expression.

\subsection{Fano formula for conductance}

For the sake of clarity we provide the Fano expression for the
conductance which can be derived from Eqs.~(\ref{conduct})-(\ref{gammaAB}) (details can be found in Ref.
[\onlinecite{Ujsaghy}]). First of all, the Green's function of the
3d orbital is assumed to be known and to have the simple form
\begin{eqnarray}
G_{dd}(\omega) &=& { A_d \over \omega - \epsilon_F - \epsilon_d - i
2 \Delta}+ {A_U \over \omega -\epsilon_F - \epsilon_d -U -i
2 \Delta}
\nonumber \\
&+& { A_K \over \omega  -\epsilon_F - \epsilon_K -i T_K}.
\label{gdd}
\end{eqnarray}
The three terms appearing in Eq.~(\ref{gdd}) correspond to the singly
occupied atomic d level situated at $\epsilon_d$, the Kondo peak
situated at $\epsilon_K$ and the doubly occupied level,
$\epsilon_d +U$ refered to the Fermi level, $\epsilon_F$.  $\Delta$ is defined as
the half-width of the
3d impurity due to its hybridizaton with the metal surface:
$\Delta(\omega)=
\pi \sum_{\bf k} |V_{\bf \tilde{k}}|^2
\delta(\omega-\epsilon_k)$.  $A_K$, $A_d$, and $A_U$ are 
spectral weights associated with the Kondo, the singly occupied and doubly occupied
adsorbate,
respectively.

Greens function $G_{dd}(\omega)$ given by Eq.~(\ref{gdd}) describes
a magnetic impurity coupled to a metallic host in the Kondo regime.
In the next section, we will give the parameters relevant to the
3d impurities in noble metals used to model $G_{dd}(\omega)$.

We introduce Eq.~(\ref{gdd}) in Eq.~(\ref{gammaAB}) and define
$\tilde{\epsilon}\equiv {\omega-\epsilon_K \over T_K}$. For
energies, $\omega < T_K$ and in the Kondo regime, $T<<T_K$,
the following Fano expression for the conductance is obtained:
\begin{equation}
G(\omega)= C(\omega) {(q+\tilde{\epsilon})^2 \over 1 +
\tilde{\epsilon}^2 }+ D(\omega),
\label{fano}
\end{equation}
where $C(\omega)$ and $D(\omega)$ have a weak dependence on $\omega$ and
$q$ denotes the Fano parameter given by Eq.~(\ref{fanoparam}).
Hence, the shape of the conductance observed, $G(\omega)$, in the
STM is governed by the value of the Fano parameter, $q$. Fig. \ref{fig1}
explains qualitatively the relation between $B(\omega)$ and conductance
lineshapes.

To conclude this section, we emphasize that Fano lineshapes
arising in the present model are a consequence solely of the
interference between metal and adsorbate waves. No direct coupling
between the tip and the 3d adsorbate is taken into account (which
in any case should be very small) nor needed to explain Fano
phenomena. However, we will see below that the tip has, in fact,
an important influence on the final shape of $ B(\omega)$ (and
$\delta \Gamma(\omega)$) through the matrix elements, $M_{\bf
\tilde{k}}$.

\section{Discussion of parameters and modelling of the system}
\label{secdis}

From previous section it becomes clear that the shape of the
conductance as a function of the bias measured by the STM depends on
the parameters modelling the adsorbate-substrate-tip interaction entering $B
(\omega)$.
Here we give details of how these parameters are obtained.

{\it A. Surface potential and metal wavefunctions}

A realistic and at the same time simple description
of $V_M$ is attained by using a JJJ potential.\cite{Jones}  This potential
interpolates
between the image potential at long distances and the
potential inside the bulk.  It provides a realistic description
of the surface barrier potential being particularly useful in interpreting
LEED and photoemission data of noble metal surfaces.\cite{Jones,Smith,Jennings}

\begin{figure}
\begin{center}
\epsfig{file=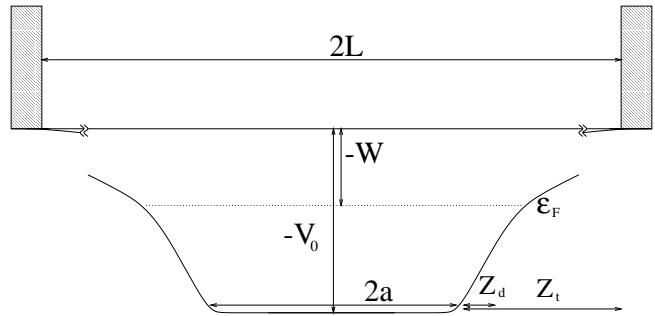,width=9.0cm,angle=0}
\end{center}
\caption{Schematic representation of the potential used to model
the noble metal surface.  The sizes of the large and small boxes
used for quantization of the wavefunctions are given by $2L$ and
$2a$, respectively.  The hatched vertical rectangles denote walls
of infinite potential. $Z_d$ and $Z_t$ denote the positions of the
adsorbate and the tip with respect to the last plane of ions.
$V_0$ is the height of the surface barrier and $W$ the metal
workfunction. The Fermi energy is referred to the vacuum level and
is denoted by the horizontal dotted line: $\epsilon_F=-W$}
\label{fig2}
\end{figure}

The JJJ potential contains three parameters that can be obtained from
fitting the potential to density functional calculations or experimental data.
The surface potential is given (in Rydberg energy units) by:
\begin{equation}
V_M(Z)=\left\{\begin{array}{ll} {-1 \over 2 (Z-Z_{im})}
(1-e^{-\lambda (Z-Z_{im})})&,\;   Z > Z_{im},
\\
\\
{-V_0 \over A e^{\beta (Z-Z_{im}) }+1 }&,\;  Z < Z_{im}
\end{array}\right.
\label{metpot}
\end{equation}
Here $V_0$ is the depth of the bulk potential and $\lambda$
controls the sharpness of the surface barrier potential. The
parameters $A=2 V_0/\lambda -1$ and $\beta=V_0/A$ are obtained by
imposing the condition of continuity of the potential at
$Z=Z_{im}$ (see Ref. [\onlinecite{Jones}] for more details).
Throughout the paper we refer the tip, $Z_t$, the adsorbate
position, $Z_d$, and the image plane position, $Z_{im}$, to the
last plane of ions situated at $a$ (see Fig. \ref{fig2}) and all
energies are referred to the vacuum level. The sharpness of the
potential barrier experienced by the electrons inside the metal is
controlled by $\lambda$.  For increasing $\lambda $, for instance, the
potential becomes gradually sharper. As we will see below parameter $\lambda$
turns out to play an important role in our model.

Metallic wavefunctions are obtained considering box quantization.
We define a large box of side $2 a$,
which describes the substrate enclosed by a larger box of side $2
L$, where the relative sizes of the boxes satisfy: $L>>a$ (see Fig. \ref{fig2}).
In practice it is sufficient to take $a/L = 0.3$.
The perpendicular dependence of the wavefunction is
obtained from the integration of Schr\"odingers equation in the
presence of the JJJ potential starting at distances of about $25$ \AA~outside the surface towards
 the metal.  Imposing periodic boundary conditions
in the
parallel direction, the crystal wavefunction reads:
\begin{equation}
\Psi({\bf r})= {1 \over \sqrt{\Omega}} e^{i \bf{k_{||} r_{||}}} \Psi_{k_z} (z)
\label{wavefun}
\end{equation}
where,  $\Psi_{k_z}(z)$, is the perpendicular part of the full
wavefunction. Sates above the vacuum are normalized to the volume
$\Omega=L^3$ while states below it are normalized to
$\Omega=a^3$.

{\it B. Tip and adsorbate wavefunctions}

The adsorbate wavefunction is described using a Slater-type
function:
\begin{equation}
\phi_i( {\bf r})=N_i r^{n-1} e^{-\eta_i r} Y_{lm}(\theta,\phi),
\label{slater}
\end{equation}
where $n$ is the main quantum number, $l$ the angular momentum and
$m$ the z-component of the angular momentum. The normalization
constant of orbital $i$ is $N_i=[(2n)!]^{-1/2}(2
\eta_i)^{(n+1/2)}$ and $Y_{lm}(\theta,\phi)$ is a conventional
spherical harmonic. For the 3d transition metal atom we use $n=3,
l=2, m=0$ with an exponential decay $\eta_d=2.3 a_0^{-1}$, where
$a_0$ is the Bohr radius.

Following Tersoff and Hamann,\cite{Tersoff} the tip wavefunction,
$|t>$, is described by a simple s-type orbital ($n=1$), with an
exponential decay fixed by the workfunction of the metal surface:
$\eta_t=\sqrt{2m_eW/\hbar^2}$, where $m_e$ is the electron mass
and $W$ is the workfunction of the metal surface which we take to
be $W = 4.95$ eV in our calculations. This choice for $\eta_t$
relies on the fact that, typically, tungsten tips are dipped in
the metal surface before measuring the conductance between the STM
tip and the surface.\cite{Schneider} Hence, it is reasonable to
describe the wavefunction tails sticking out from the tip with the
same exponential decay as the ones describing the metal surface.

{\it C. Computation of matrix elements}

Once we have obtained the wavefunctions of the
substrate we may proceed with the computation
of the hybridization and overlap matrix elements defined in Eq.~(\ref{matrixel}).
For simplicity we
discuss the adsorbate-substrate hybridization matrix elements.
From Eq.~(\ref{wavefun}) hybridization
and overlap matrix elements expressed in cylindrical coordinates, read:
\begin{eqnarray}
V_{\bf k} &=& 2 \pi \int_0^{\infty} d r_{||} r_{||}
\int_0^{\infty} d z
\Psi_{k_z}(z) J_0(k_{||} r_{||}) V_M(z) \phi_d(|\bf{r-r_d}|)
\nonumber \\
S_{\bf k} &=& 2 \pi \int_0^{\infty} d r_{||} r_{||}
\int_0^{\infty} d z
\Psi_{k_z}(z) J_0(k_{||} r_{||}) \phi_d(|\bf{r-r_d}|)
\label{matrix}
\end{eqnarray}
where $J_0(k_{||} r_{||})$ is the cylindrical Bessel function of
zeroth order, $V_M$, is the surface potential and $\phi_d(|{\bf
r - r_d}|)$ is the adsorbate wavefunction. The computation of the
tip-substrate hybridization matrix elements, $M_{\bf k}$, proceeds
along similar lines replacing $\phi_d(|{\bf r - r_d}|)$ by
$\phi_t(|{\bf r - R}|)$.

\begin{figure}
\begin{center}
\epsfig{file=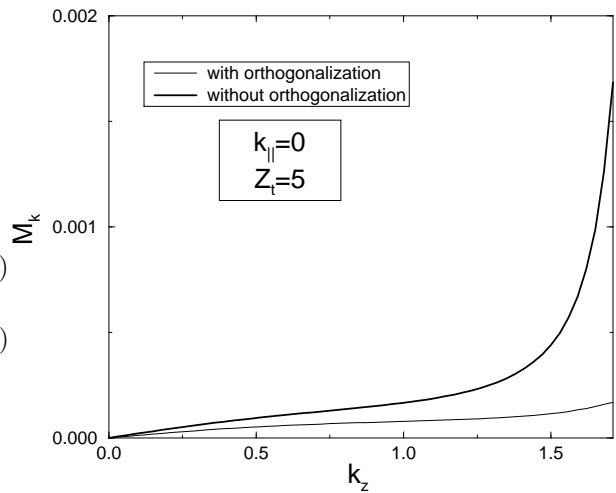,width=9.0cm,angle=0}
\end{center}
\caption{Effect of orthogonalization on tip-substrate hybridization matrix elements. 
While hybridization matrix elements computed between the original substrate wavefunctions 
and the tip depend strongly on perpendicular momentum, $k_z$, 
this dependence is weaker for matrix elements containing orthogonalization
corrections.  The JJJ potential used in this calculation is parameterized using 
$\lambda=1.7$ \AA$^{-1}$, $V_0 =13.05$ eV, and
$Z_{im}=1.1$ \AA.  The tip is located at $Z_t=5$ \AA~above
the surface plane of ions on top of the adsorbate ($R_{||}=0$). $k_z$ is given
in \AA$^{-1}$ with the vacuum level corresponding to $k_0=1.85$ \AA$^{-1}$.  
}
\label{fig3}
\end{figure}

The behavior of tip matrix elements deserves careful attention.
Due to the slow decay of the tip wave function, most of the contribution
to $M_{\bf \tilde{k}}$ comes from the region close to the metal surface for 
$k_z$ not too large. Matrix elements $M_{\bf \tilde{k}}({\bf R})$ can then be 
approximately factorized as\cite{Plihal}:
\begin{equation}
M_{\bf \tilde{k}}({\bf R}) \approx f(Z_t) e^{i {\bf k_{||}} {\bf R_{||}} } 
\tilde{M}_{\bf \tilde{k}}.
\label{matrixtip}
\end{equation}
where $\tilde{M}_{\bf \tilde{k}}$ is independent of the
tip position.  This remains true for $k_z \lesssim 1.6-1.7$ \AA$^{-1}$, which is 
close to $k_0=1.85$ \AA$^{-1}$ (for $V_0=13.05$ eV). The function $f(Z_t)$
is related to the value of the tip wave function in the surface region and
it decays rapidly\cite{matrixtip} with $Z_t$.  $\tilde{M}_{\bf \tilde{k}}$ is 
found to have a weak dependence on $k_z$ in contrast to matrix elements 
obtained from the original substrate wavefunctions as can be observed in 
Fig.  \ref{fig3}. Indeed, using $V_M$ in the matrix elements as imposed by the orthogonalization
cuts off part of the exponential dependence of metal wavefunctions far from the surface
leading to a weaker dependence on $k_z$.  This is crucial to obtain reasonable
shapes of $B(\omega)$.

In order to have a qualitative understanding of the behavior of
adsorbate-substrate hybridization matrix elements, $V_{\bf k}$, we focus on
electrons which have no parallel momentum, $k_{||}=0$. Matrix
elements gradually increase with increasing $k_z$ due to
increasing overlap of the metal states with
the adsorbate.  The behavior of matrix elements at large $k_z$ can be
partially understood by analyzing Eq.~(\ref{matrix}) with
$\Psi_{k_z}(z)$ nearly constant, (which is the case for the metal wavefunction
outside the surface).   For 
a $d_{3z^2-r^2}$ adsorbate the associated matrix elements, $V_{\bf \tilde{k}}$ 
would tend to zero due to cancellations associated with the lobes of the
3d-orbital.  Indeed, we find that while $V_{\bf \tilde{k}}$ computed  for a 
d-orbital presents a downturn at intermediate $k_z \approx 1.4$ \AA$^{-1}$,  $V_{\bf \tilde{k}}$ 
computed with an s-type orbital increases rapidly up to $k_z \approx k_0$. Orthogonalization effects are also important as
can be noticed from the fact that $V_{\bf k} \approx <d|V_M|d>
S_{\bf k}$ as $k_z \rightarrow k_0$.  Hence, we find that
matrix elements computed using orthogonalized wavefunctions, $V_{\bf \tilde{k}}$,
initially increase with $k_z$, reach a maximum and then 
they are gradually suppressed regardless of the angular dependence of the adsorbate
orbital.  This dependence of 
$V_{\bf \tilde{k}}$ is reflected in the frequency dependence of $B(\omega)$. 
The above discussion illustrates the importance of appropriately including 
orthogonalization effects in the computation of hybridization matrix elements. 

The behavior of matrix elements at very large perpendicular momentum
needs to be carefully examined.  Metallic waves with large
momentum (corresponding to energies far above the vacuum level),
contain rapid oscillations which lead to cancellations in the
integrand of Eq.~(\ref{matrix}). A high energy cut-off may be simply
estimated from: $k^{cutoff}_i \sim \eta_i$.
This gives $\omega^{cutoff}_d \approx 70$ eV for the adsorbate and
$\omega^{cutoff}_t \approx 5$ eV for the tip. Hence,
hybridization matrix elements and in turn $B(\omega)$ will be
non-zero in the energy range $-V_0 <  \omega < \omega^{cutoff}_t$.
It is remarkable that the tip cuts off all high energy
contributions having an important influence on the final shape of
$B(\omega)$ and in turn on conductance lineshapes.  This point
seems to have been overlooked before.

{\it D. Computation of $B(\omega)$ }

Once we have the hybridization matrix elements, it is
straightforward, following Ref. [\onlinecite{Plihal}] to
obtain the energy dependence of $B(\omega)$
given in Eq.~(\ref{B}).  After performing integration
in parallel momentum, $B(\omega)$ can be written as
\begin{equation}
B(\omega)= {\Omega \over \pi} \sqrt{ {2 m_e^3  \over \hbar^6}
(\omega+ V_0) } \int_0^1 d x V_{\tilde{k}_{||}, \tilde{k}_z}
M_{\tilde{k}_{||}, \tilde{k}_z},
\end{equation}
where $\tilde{k}_{||}=\sqrt{2 m_e/ \hbar^2 (\omega+V_0)} \sqrt{1-x^2}$ and
$\tilde{k}_z=\sqrt{2 m_e/ \hbar^2 (\omega+V_0)} x$ and $\Omega$ is the
volume. Note the factor $\Omega$ appearing in front of the integral
cancels out the volume coming from the normalization of the
surface wavefunctions coming in the matrix elements.

{\it E. Parametrization of adsorbate Greens function}

In order to compute the modification induced by the adsorbate on
conductance lineshapes, $\delta \Gamma (\omega)$, knowledge of
$G_{dd}(\omega)$ is needed. We take parameters used by Ujsaghy
{\it et. al.} \cite{Ujsaghy} to model Co on Au: $\epsilon_K=3$
meV, $\epsilon_d=-0.84$ eV, $U=2.8$ eV,  $A_K=\pi T_K/\Delta$,
$A_d=0.45$ and $A_U=1-A_d-A_K$ with the Kondo temperature fixed
at $T_K=5$ meV. As we are particularly interested in what is
happening close to the Fermi energy (at energy scales of the order
of $T_K$) the relevant parameters to our calculations are $T_K$
and $\epsilon_K$. Hence, these are input parameters which could be
obtained from other sources such as experimental data or {\it ab
initio} density functional calculations. It is not the purpose of
the present work to have a first-principles determination of Kondo
temperatures for different adsorbate/substrate systems but rather
to  have a qualitative understanding of conductance lineshapes.

\section{Discussion of results}
\label{secres} In this section we show our main results and
discuss their relevance to experiments.  A positive Fano parameter
$q \geq 0$ is typically found in our calculations for parameters
charaterizing the noble metal surface, the adsorbate and their
mutual interaction.  We explore how lineshapes depend on the
parameters characterizing the atom-surface interaction. These
parameters are:  the position of the tip  ${\bf R} = ({\bf
R_{||}}, Z_t)$, the adsorbate-substrate distance, $Z_d$, 
and the shape of the surface potential controlled by $\lambda$.  As there is cylindrical symmetry in
our model,  our results only depend on the absolute magnitude of
the tip-adsorbate lateral displacement $R_{||}=|{\bf R_{||}}|$.
We do not find much dependence of our results with the
workfunction or the height of the surface barrier so we fix $V_0=13.05$ eV
and the workfunction $W=4.95$ eV, which are typical values for
noble metal surfaces.  We will first discuss the dependence of
lineshapes with the tip fixed on top of the adsorbate, $R_{||}=0$
at a distance of about $Z_t=5$ \AA.  At the end of the section we
will study how lineshapes depend on $R_{||}$ and $Z_t$.

\subsection{Dependence of Fano lineshapes with adsorbate-surface distance}

\begin{figure}
\begin{center}
\epsfig{file=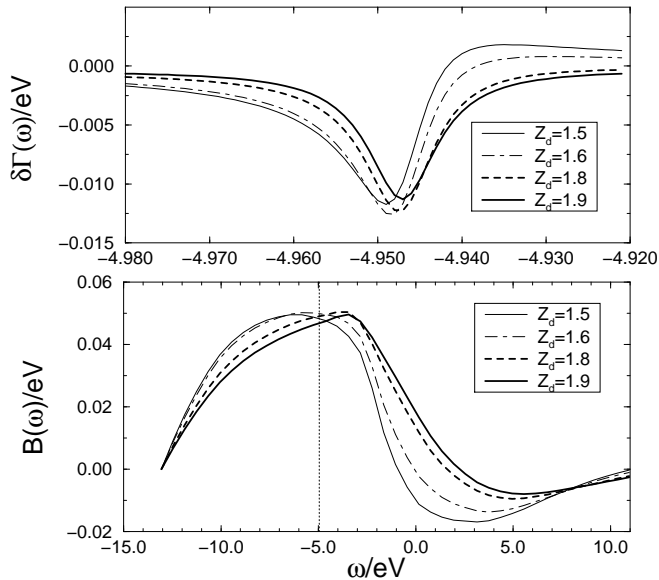,width=9.0cm,angle=0}
\end{center}
\caption{Dependence of lineshapes on the adsorbate-metal
adsorption distance. As the adsorbate is brought closer to
the surface the lineshapes become gradually more asymmetric.
The lower pannel displays the function $B(\omega)$ and
in the upper one is the corresponding modification in the
tip-surface interaction induced by the adsorbate: $\delta \Gamma (\omega)$.
The JJJ potential is parametrized using $\lambda=2.08$ \AA$^{-1}$ and
$Z_{im}=1.1$ \AA.  The tip is located at $Z_t=5$ \AA~above
the surface plane of ions.  All distances in the figure
are given in \AA~and refered to the last plane of ions.
The vertical dotted line denotes the Fermi level position:
$\epsilon_F=-W=-4.95$ eV. The height of the surface barrier is $V_0=13.05$ eV. }
\label{fig4}
\end{figure}

Fixing the tip at $R_{||}=0$ (on top of the adsorbate) and $Z_t=5$ \AA~we 
analyze how lineshapes depend on the adsorbtion distance,
$Z_d$. In Fig. \ref{fig4} the dependence of $B(\omega)$ and
$\delta \Gamma(\omega)$ is shown for typical parameters describing
noble metal surfaces: $\lambda \approx 2.08$ \AA$^{-1}$, $V_0
\approx 13.05$ eV at typical adsorbate-surface distances: $Z_d \approx
1.5-2$ \AA. Adsorbtion distances for different surfaces are
obtained assuming hard spheres for the atoms.  At large adsorbtion
distances, $Z_d \approx 2$ \AA, $B(\omega)$ is found to be rather
symmetric with respect to the Fermi energy and consequently (see
Fig. 2) the lineshape is a nearly symmetric dip.  As the adsorbate
is moved closer to the surface ($Z_d$ decreasing) the frequency
dependence of $B(\omega)$ varies and consequently the lineshapes
too.  An enhancement of weight below the Fermi energy as well as a
sharper drop above it occurs.  This leads to a variation of
lineshapes from $q \approx 0$ at large distances towards $q
\gtrsim 0$ at short adsorbate-surface distances.

To understand these results, we notice that substrate states with low
energy have a strong exponential decay outside the surface. As the adsorbate
is brought closer to the surface, the overlap of the adsorbate orbital with these
states increases. This leads to an increase of $B(\omega)$ at small $\omega$.  States with
a larger energy extend farther outside the surface and have a substantial overlap
with the different lobes of the adsorbate $3d$-orbital.  This tends to lead to
a cancellation of the different contributions to $V_{\bf k}$.  This tendency
increases as the adsorbate is brought closer to the surface, explaining the
reduction of $B(\omega)$ at larger $\omega$.

\begin{figure}
\begin{center}
\epsfig{file=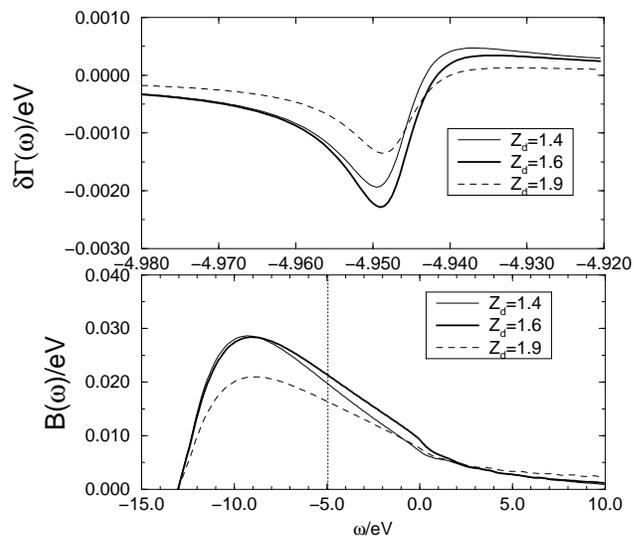,width=9.0cm,angle=0}
\end{center}
\caption{Dependence of lineshapes on the adsorbate-metal
adsorption distance for a Jellium model with a step potential.
In contrast to the JJJ model potential (see Fig. \ref{fig4}) there is a weak
dependence of lineshapes with the adsorbate-substrate distance.
The tip is located at $Z_t=5$ \AA~above the surface plane of ions.
All distances in the figure
are given in \AA. The vertical dotted line denotes the Fermi level position:
$\epsilon_F=-W=-4.95$ eV. The height of the surface barrier is $V_0=13.05$ eV. }
\label{fig5}
\end{figure}

For comparison we also show in Fig. \ref{fig5} the dependence of
lineshapes with the adsorbate-surface distance for a Jellium with
a step potential. Lineshapes are found to be
somewhat asymmetric as a large part of the weight
in situated at low energies in $B(\omega)$  (see also Fig.
\ref{fig1}).  A weaker dependence of $B(\omega)$ with the
adsorbate-surface distance is obtained
as compared to the JJJ potential. 
The fact that we obtain reasonable lineshapes with a simple model
such as a Jellium with a step potential can be attributed to 
orthogonalization effects.  Tip-substrate matrix elements computed without orthogonalization 
corrections would increase rapidly with $k_z$ as they involve integrations
over the whole space (instead of the region close to the surface) 
leading to rather asymmetric shapes of $B(\omega)$.

\subsection{Dependence of lineshapes with the shape of the potential barrier}

\begin{figure}
\begin{center}
\epsfig{file=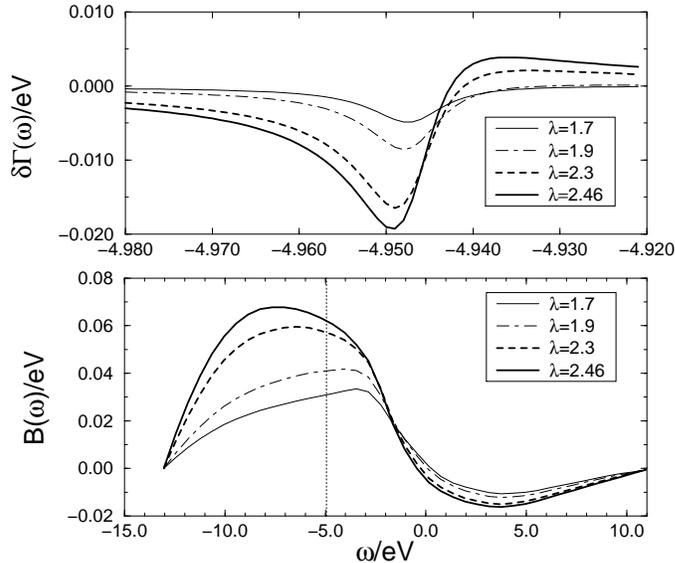,width=9.0cm,angle=0}
\end{center}
\caption{Dependence of lineshapes with the sharpness of the
surface potential.  As $\lambda$ increases $B(\omega)$ and lineshapes
become more asymmetric. The lower pannel shows $B(\omega)$ for different
$\lambda$ and the upper one the corresponding lineshapes.  The surface
potential is
modelled for different $\lambda$ and $Z_{im}=1.1$ \AA.  The position of the
tip and adsorbate are $Z_t=5$ \AA~and $Z_d=1.6$ \AA,  respectively.
All distances are given in units of \AA~and $\lambda$ is in units of \AA$^{-1}
$.
The vertical dotted line denotes the position of the Fermi level: $\epsilon_F=-
W=-4.95$ eV.}
\label{fig6}
\end{figure}

There are two parameters that can be tuned to change the shape of
the surface potential. One is the image plane position, $Z_{im}$,
and the other is the sharpness of the surface potential given by
$\lambda$. We fix $Z_{im}=1.1$ \AA~and study the dependence of
$B(\omega)$ and lineshapes $\delta \Gamma (\omega)$, with the
sharpness of the potential barrier.  This dependence is shown in
Fig. \ref{fig6}. We find that an increase in $\lambda$ leads to a
more asymmetric $B(\omega)$ and the associated conductance
lineshape, $\delta \Gamma(\omega)$ with more positive values of
the Fano parameter, $q$.

Increasing $\lambda$ makes the metal potential, $V_M$, sharper
so that the exponential tails of the metal wavefunctions outside
the surface are shifted closer to the adsorbate leading to a
stronger metal-adsorbate overlap. This is analogous to bringing
the adsorbate closer to the surface for a fixed $\lambda$ as
discussed previously. Hence, an increase in $\lambda$ leads to an
increase in the weight of $B(\omega)$ at low energies and a
sharper drop at energies above the Fermi level. The stronger
dependence found for $B(\omega)$ with $\lambda$ than with the
adsorbate-substrate distance can be ascribed to the exponential
dependence of $V_M$ with $\lambda$.

Summarizing, we find that lineshapes are strongly dependent on the
degree of sharpness of the metal potential. Making the surface
potential sharper leads to more asymmetric lineshapes at typical
atom-substrate adsorption distances.

\subsection{Dependence of conductance lineshapes with tip position}

\begin{figure}
\begin{center}
\epsfig{file=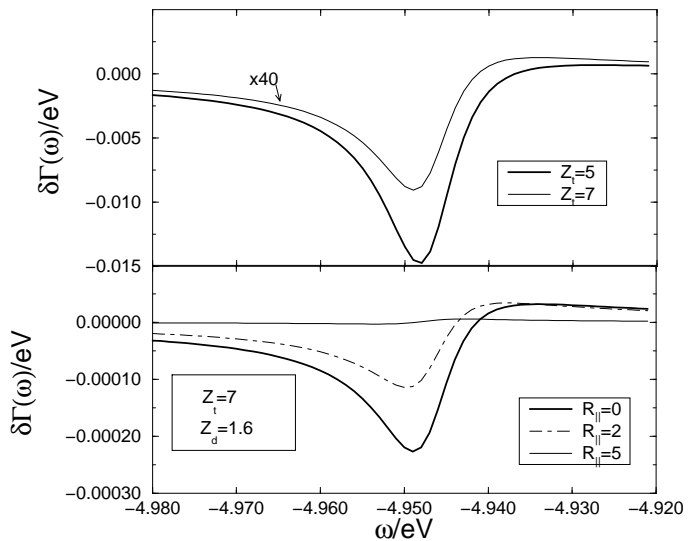,width=9.0cm,angle=0}
\end{center}
\caption{ Dependence of lineshapes on the position of the tip.
The upper panel shows the small perpendicular dependence of
lineshapes if the tip is moved outwards from the surface by 2~\AA. The surface potential is parametrized by $\lambda=2.1$ \AA$^{-1}$
and the position of the image plane $Z_{im}=1.1$ \AA. The
adsorbate is at $Z_d=1.6$ \AA.  All distances in the plot are given in \AA~and the Fermi
level is situated at: $\epsilon_F=-4.95$ eV. The conductance
lineshape at $Z_t=7$ \AA~is multiplied by 40 for comparison with
$Z_t=5$ \AA.} \label{fig7}
\end{figure}

We finally discuss the dependence of conductance lineshapes when
the tip is moved away from the adsorbate either laterally ($R_{||}
\neq 0$) or perpendicularly.  The tip position comes into our
calculations through matrix elements, $M_{\bf \tilde{k}}$. A weak
dependence of lineshapes with perpendicular tip-substrate
distance, $Z_t$ is found. Varying the distance from 5 to 7 \AA~leads 
to a small change in the lineshape becoming slightly more
asymmetric (see Fig. \ref{fig7}).  The weak dependence of the line shape
on the tip-surface distance can be understood from Eq.~(\ref{matrixtip}).
This equation predicts a strong reduction of the amplitude of the 
conductance with vertical displacement of the tip $\sim f(Z_t)^2$,
but no change in its shape.  The weak dependence found is consistent 
with experimental observations which also indicate that the
direct tip-adsorbate coupling is very small (as expected from the localized
nature of the d-orbital of the adsorbate).

We can also analyze how lineshapes depend on the lateral position
of the tip, $R_{||}$. An example is shown in Fig. \ref{fig7}
which a rapid decay of the amplitude of the lineshapes is found.
Lineshapes are found to depend slightly on $R_{||}$ at tip-surface
distances of about $Z_t= 7$ \AA~which is the typical experimental
tip position.  Indeed, crystal wavefunctions with $k_{||}=0$ have
the largest amplitude outside the surface representing electrons
which have the largest probability of being detected by the tip
(see Ref. [\onlinecite{Plihal}] for discussion). Hence, the
oscillatory behavior appearing in Eq.~(\ref{matrixtip}) is
suppressed as only the states with small $k_{||}$ contribute to
the momentum sums.

The rapid decay of the amplitude with $R_{||}$ obtained from our
calculations is consistent with the disappearance of the features
in the conductance at distances of about $R_{||} \sim 5-10$ \AA.
However, the small variation of the shape of the conductance with
$R_{||}$ found is only partially consistent with experiments as
for some systems such as Co on Cu(100) or Co on Au(111) lineshapes
become gradually more symmetric ($q \rightarrow 0$) at larger
$R_{||}$.  It is worth mentioning that in the case of Co on
Au(111), the lateral dependence of lineshapes depends not only on
the modulus of ${\bf R_{||}}$ but also on the {\it direction} in
which the tip is moved \cite{Madhavan1} from the adsorbate. This
cannot be described in the present framework as the surface
corrugation is not included in our model.

The finite size of the tip leads to a
suppression of the tip-substrate matrix elements at large energies
due to rapid oscillations of the metal wavefunctions providing a
high energy cut-off. This cut-off turns out to be crucial to
obtain reasonable lineshapes. Had we used a point-like tip the
amplitude of $B(\omega)$ at energies below the vacuum level would
be exponentially suppressed while above it large oscillations
would occur.  This is because wavefunctions are evaluated far from
the surface (at the tip position). Hence, for a point-like tip the
Fano parameter would be very large ($q>>1$) and a peak rather than
a dip would appear in conductance measurements. 

\section{An application to conductance measurements of Cobalt on
Copper surfaces}
\label{secco}
In the previous sections a detailed description of the method used has
been provided.  In the following we provide the reader with an application
to conductance measurements of Cu surfaces with Co atoms deposited on them.

\begin{figure}
\begin{center}
\epsfig{file=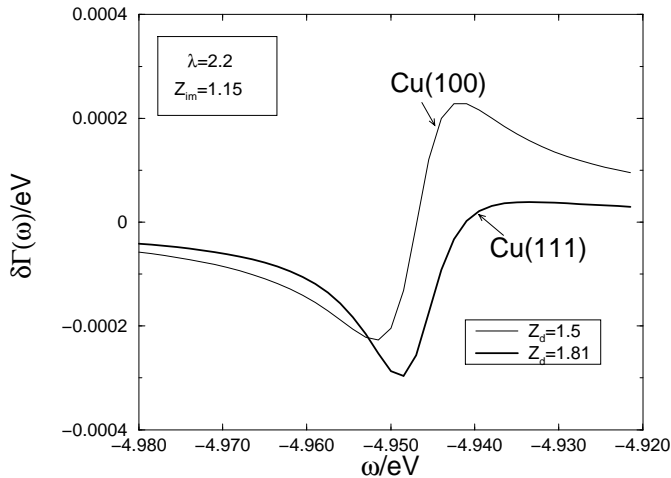,width=9.0cm,angle=0}
\end{center}
\caption{Lineshapes associated with conductance measurements of Co on Cu(100)
and Cu(111).
Parameters taken from Ref. [\onlinecite{Smith}] are used to
describe the surface potential:  $\lambda=2.2$ \AA$^{-1}$ and $Z_{im}=1.15$ \AA.
The adsorbtion distance is larger for Cu(111) than Cu(100) by about $0.3$ \AA.
This leads to a more asymmetric lineshape in the latter than in the former
case in agreement with experimental findings.\cite{Knorr}  All distances in
the figure are given in \AA~and $\lambda$ is given in \AA$^{-1}$.
}
\label{fig8}
\end{figure}

Copper surfaces are described using the parameters quoted in Ref.
[\onlinecite{Smith}] for the analysis of surface states in noble
metal surfaces.  For Cu surfaces we take $\lambda=2.2$ \AA$^{-1}$
and the adsorbate -substrate distance is estimated assuming
hard spheres for the atoms.\cite{Pauling}  As the (111) surface is more closely
packed than the (100) surface the Co atom adsorbs at larger
distances in the former than in the latter surface. This leads to
$Z_d=1.5$ \AA~and $Z_d=1.8$ \AA~for the (100) and (111) surfaces,
respectively.  At the temperatures used in the experimental studies of the 
conductance,\cite{Knorr} Co sits outside the Cu(100) surface and it is not 
incorporated in the surface, as happens at higher temperatures \cite{Levanov}. 
Ab initio calculations \cite{Scheffler} then give $Z_d=1.5$ \AA, in agreement
with our simple estimate.
We find that difference in the adsorbtion distance is enough to produce changes 
in the lineshapes as can be observed in Fig. \ref{fig8}. From the dependence of
lineshapes with $Z_d$ discussed above we find that for Co on
Cu(100) surfaces lineshapes would be more asymmetric than for Co
on Cu(111).  This is consistent with experiments performed for 
these two different surfaces.\cite{Knorr,LNJ}  Our result may be of
more general validity and may apply to conductance measurements on
other metal surfaces such as Ag or Au. However, care should be
taken as our model contains the adsorbate-substrate distance only:
corrugation effects may also be important.  More experiments that
systematically analyze the dependence of lineshapes for Co on
different crystal faces are needed to corroborate our findings.

\section{Conclusions}

We have introduced an Anderson model in an appropriate
orthogonalized basis to analyze STM measurements of noble metal
surfaces with adsorbed transition metal atoms. A realistic surface
potential (JJJ potential), a single $d_{3z^2-r^2}$ adsorbate
orbital and a s-tip orbital are used to compute the parameters of
the model. For typical values of the parameters characterizing the
adsorbate-substrate interaction, we find lineshapes with Fano
parameters, $q \geq 0$, in agreement with experimental trends.

We have introduced a function $B(\omega)$ which describes the
adsorbate-substrate interaction. The description of experimental
observations requires a $B(\omega)$ which is fairly symmetric with
respect to the Fermi energy. Since the underlying density of
states is approximately parabolic and the matrix elements tend to
further reduce $B(\omega)$ at small $\omega$, it is interesting
that the model, nevertheless, gives reasonable shapes for
$B(\omega)$. We find that the tip matrix elements play an
important role. The tip wave function is rather extended, and
matrix elements to continuum states at high energies are very
small, due to the rapid oscillations of the continuum states 
over the spatial range of the tip wave function. This helps to make $B(\omega)$ more symmetric around
$\omega=\epsilon_F$.  We also find that it is essential to use substrate 
states which have been orthogonalized to the adsorbate and the tip.  Even when the realistic JJJ potential is
used, lineshapes are found to strongly depend on the sharpness of
the surface potential controlled by $\lambda$ becoming more
asymmetric with increasing $\lambda$ ({\it i. e.} for sharper
potentials). To a lower degree conductance lineshapes are also
found to depend on the adsorbtion distance: the closer to the
surface the adsorbate is the more asymmetric the lineshapes
become. These tendencies can be explained from the enhancement in
the coupling of the metal wavefunctions and the
adsorbate at low energies together with the suppression at energies above the Fermi
level induced by the angular dependence of the $d_{3z^2-r^2}$
orbital and the orthogonalization.   

  More closely packed surfaces
such as Cu(111) are expected to lead to more symmetric lineshapes
than the more open Cu(100) faces as Co adsorbs further out in the former than in the
latter.  We find good agreement with
experiments comparing lineshapes of Co on Cu(100) with Co on
Cu(111) surfaces.
In agreement with experiments we also find that
the tip-substrate distance does not strongly influence the shape
of the conductance.

\acknowledgments
We acknowledge helpful discussions about experiments with M. A.
Schneider and P. Wahl and for pointing out to us several references.
J. M. was supported by a Marie Curie Fellowship
of the European Community program "Improving Human Potential" under
contract No. HPMF-CT-2000-00870.

\end{document}